\documentclass{elsart}
\usepackage{graphicx}

\begin{document}

\begin{frontmatter}

\title{Alternative Methods to Finding Patterns in HiRes Stereo Data.} 
\author[utah]{R.U.~Abbasi\corauthref{cor1},}
\author[utah]{T.~Abu-Zayyad,}
\author[lanl]{J.F.~Amman,}
\author[utah]{G.C.~Archbold,}
\author[utah]{K.~Belov,}
\author[utah]{S.A.~Blake,}
\author[utah]{J.W.~Belz,}
\author[col]{S.~BenZvi,}
\author[rut]{D.R.~Bergman,}
\author[col]{J.H.~Boyer,}
\author[utah]{G.W.~Burt,}
\author[utah]{Z.~Cao,}
\author[col]{B.M.~Connolly,}
\author[utah]{W.~Deng,}
\author[utah]{Y.~Fedorova,}
\author[utah]{J.~Findlay,}
\author[col]{C.B.~Finley,}
\author[utah]{R.C.~Gray}
\author[utah]{W.F.~Hanlon,}
\author[lanl]{C.M.~Hoffman,}
\author[lanl]{M.H.~Holzscheiter,}
\author[rut]{G.A.~Hughes,}
\author[lanl]{P.~H\"{u}ntemeyer,}
\author[utah]{B.F~Jones}
\author[utah]{C.C.H.~Jui,}
\author[utah]{K.~Kim,}
\author[umt]{M.A.~Kirn,}
\author[col]{B.C.~Knapp,}
\author[utah]{E.C.~Loh}
\author[utah]{M.M.~Maestas,}
\author[japan]{N.~Manago,}
\author[col]{E.J.~Mannel,}
\author[lanl]{L.J.~Marek,}
\author[utah]{K.~Martins,}
\author[unm]{J.A.J.~Matthews,}
\author[utah]{J.N.~Matthews,}
\author[utah]{S.A.~Moore}
\author[col]{A.~O'Neill,}
\author[lanl]{C.A.~Painter,}
\author[rut]{L.~Perera,}
\author[utah]{K.~Reil,}
\author[utah]{R.~Riehle,}
\author[unm]{M.~Roberts,}
\author[utah]{D.~Rodriguez,}
\author[japan]{M.~Sasaki,}
\author[rut]{S.~Schnetzer,}
\author[rut]{L.M.~Scott,}
\author[col]{M.~Seman,}
\author[lanl]{G.~Sinnis,}
\author[utah]{J.D.~Smith,}
\author[utah]{P.~Sokolsky,}
\author[col]{C.~Song,}
\author[utah]{R.W.~Springer,}
\author[utah]{B.T.~Stokes,}
\author[utah]{J.R.~Thomas,}
\author[utah]{S.B.~Thomas,}
\author[rut]{G.B.~Thomson,}
\author[lanl]{D.~Tupa,}
\author[col]{S.~Westerhoff,}
\author[utah]{L.R.~Weincke,}
\author[rut]{A.~Zech,}
\author[col]{X.~Zhang,}

\address[utah]{University of Utah,
Department of Physics and High Energy Astrophysics Institute,
Salt Lake City, UT~84112, USA}
\address[lanl]{Los Alamos National Laboratory, Los Alamos, NM 87545, USA}
\address[uad]{University of Adelaide, Department of Physics, Adelaide, South Australia, Australia}
\address[umt]{Montana State University, Department of Physics, Bozeman, Montana, USA}
\address[rut]{The State University of New Jersey, Department of Physics and Astronomy, Piscataway, New Jersey, USA}
\address[col]{Columbia University, Department of Physics and Nevis Laboratory, New York, New York, USA}
\address[unm]{University of New Mexico, Department of Physics and Astronomy, Albuquerque, New Mexico, USA}
\address[japan]{University of Tokyo, Institute for Cosmic Ray Research, Kashiwa, Japan}
\corauth[cor1]{
Corresponding~author.\ {\it E-mail~address}:~rasha@cosmic.utah.edu
}

\newpage

\begin{abstract}
 In this paper Ultra High Energy Cosmic Rays UHECRs data observed by the HiRes fluorescence detector in stereo mode is analyzed to search for events in the sky with an arrival direction lying on a great circle. Such structure is known as the arc structure. The arc structure is expected when the charged cosmic rays pass through the galactic magnetic field. The arcs searched for could represent a broad or a small scale anisotropy depending on the proposed source model for the UHECRs. The Arcs in this paper are looked for using Hough transform were Hough transform is a technique used to looking for patterns in images. No statistically significant arcs were found in this study. 
\end{abstract}


\end{frontmatter}

\section{Introduction}
\label{sect-intro}
 
\begin{figure}[!hpt]
\begin{center}
\framebox{\includegraphics[width=15.0cm]{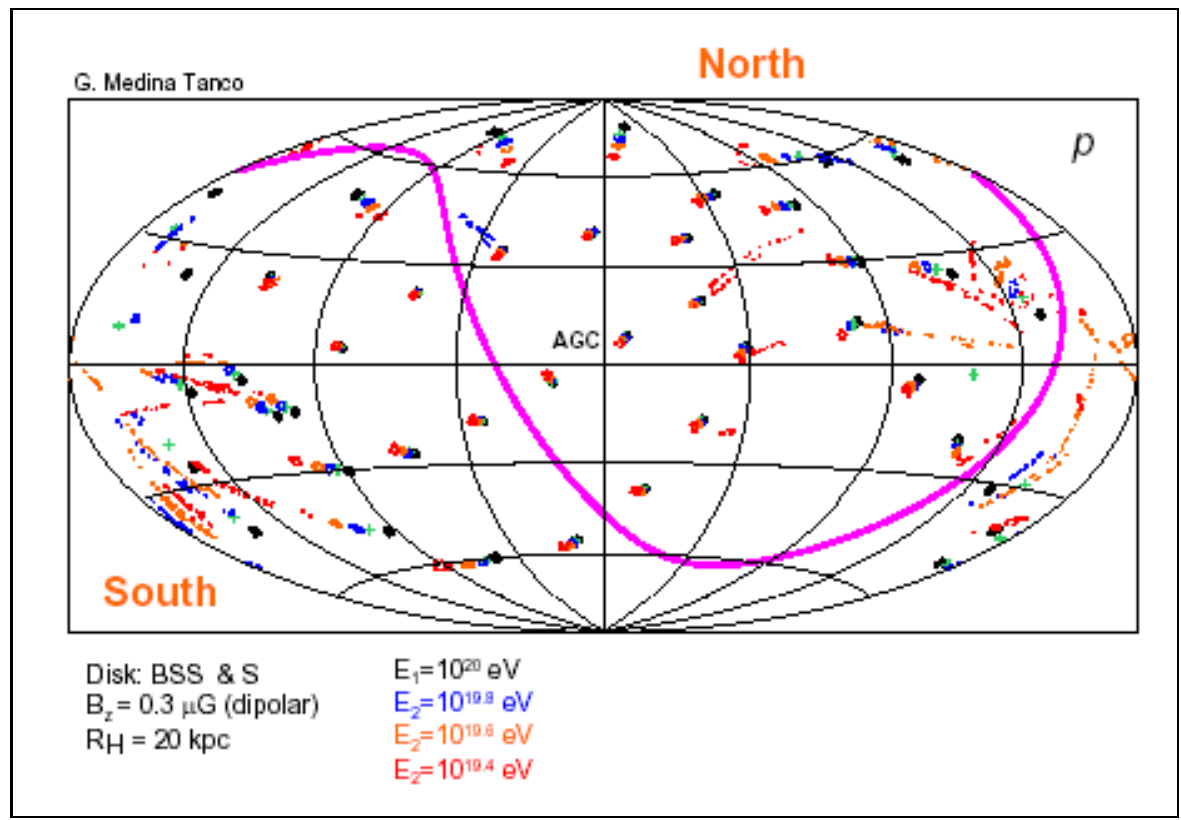}}
\caption{\label{OLENTO}Simulated ``polarization'' plot by Gustavo Medina-Tanco ~\cite{r0}. It shows cosmic rays with energies between ${10}^{19.4}$ and  ${10}^{20}$ eV from point source. The events are deflected while passing through the galactic magnetic field forming arc shapes. While the events originate at point sources, the magnetic field deflects the events to form the arc shapes which one sees in the plot.}
\end{center}
\end{figure}

Cosmic rays passing through the galactic magnetic field are deflected. The degree of deflection depends upon their energy and the strength and structure of the field. This deflection has the potential of producing either large scale or small scale anisotropy in the cosmic ray arrival direction. The scale of this anisotropy is determined by the UHECR source model. In particular, with moderate deflection through a coherent field, the pointing directions of the deflected cosmic rays from a point source may form an arc in the sky, as illustrated in Figure~\ref{OLENTO} ~\cite{r0}. Alternately, an extended candidate source, such as the Galaxy, may lead to a lengthened circle of events. These are the type of structures we would like to identify.

In this study we use the Hough transform to search for arc-like structure. Hough transform is a class of duality transformations commonly used to identify 1-dimensional structures in 2-dimensional graphical images ~\cite{r1}. The goal of this study is to look for arcs in the HiRes stereo data set. Here we use the term ``arc'' to describe a segment of great circles across the sphere of the sky.

\begin{figure}
  \includegraphics[angle=0.,width=\columnwidth]{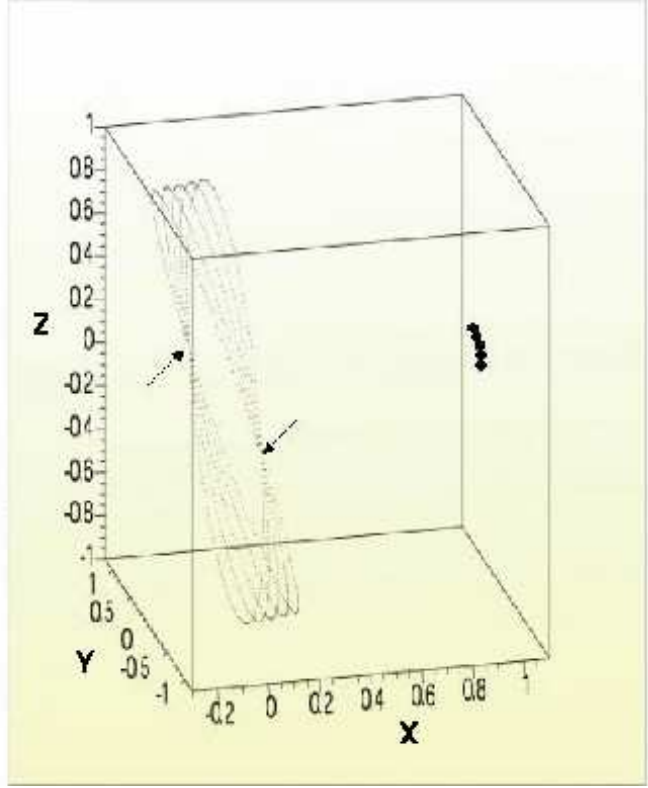}
  \caption{Illustration of the intersection of five dual-space circles associated with five events that lie on an arc. The arrows point to the two intersection points.}
  \label{CURVE}
\end{figure}

\begin{figure}[!hpt]
\begin{center}
\framebox{\includegraphics[width=9.0cm]{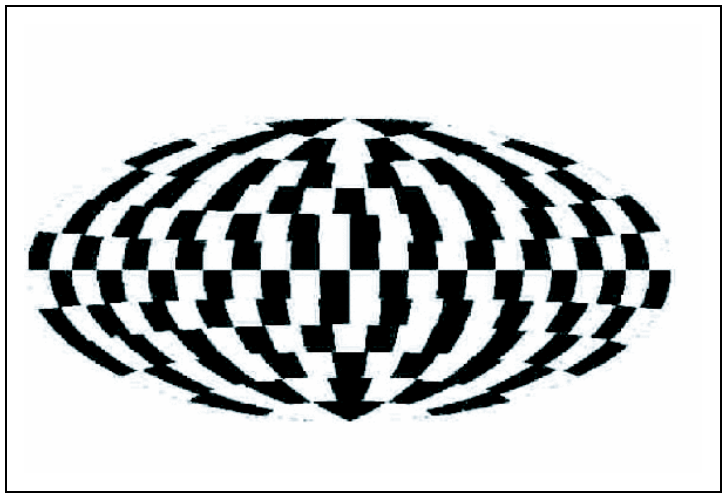}}
\caption{An illustration of the sky divided into coarse, equal-sized and equal-shaped bins. A finer bin size is used in the actual analysis ~\cite{r31}.}
\label{binning}
\end{center}
\end{figure}

In this study, the Hough transform is applied as a self-dual mapping on a sphere, where the pointing direction of each event is mapped to a great circle that is perpendicular to its arrival direction. For events that lie on an arc, their associated great circles would intersect at two points, as illustrated in Figure~\ref{CURVE}. Thus, to identify candidate arcs, we divide the dual-space for the events (which is also a sphere) into equal sized and shaped bins. A coarse example of the binning scheme ~\cite{r31} is shown in Figure~\ref{binning}. For our data analysis, we loop over all events in our data set. For each event, we increment those bins through which the corresponding great circle passes. The bin values therefore represent the total number of great circles that lie within its boundaries. The arcs are then identified by an excess in a particular bin.

To match the dual-space statistics corresponding to the 271 events in our data set ~\cite{r2}, we chose a $6^{\circ} \times 6^{\circ} $ bin-size, giving a total of 4584 bins. For display purposes (figures ~\ref{his},~\ref{sig}) the bins are numbered in latitude from north pole to south and around the axis. The dual-space bin contents are then compared to expectations from an entirely random, isotropically generated arrival distribution (without point or extended sources) to identify significant excesses.

\section{Applying The Hough Transform.}
\label{sect-detector}

We apply the Hough transform method to the stereo HiRes data to search for candidate arcs of events. The data selected are those events with energies $>{10}^{19}$eV. Particles with energies $<{10}^{19}$ eV are expected to be deflected by larger angles such that any structure would likely be lost. We exclude them as not to dilute any potential signals in the data set. This data set contains 271 events collected between  December 1999 and 31 January 2004. The event selection and quality cuts are identical to previous published study using the HiRes stereo data. Details can be found in ~\cite{r2} and ~\cite{r3}.

\begin{center}
  \begin{figure}
    \includegraphics[width=16cm]{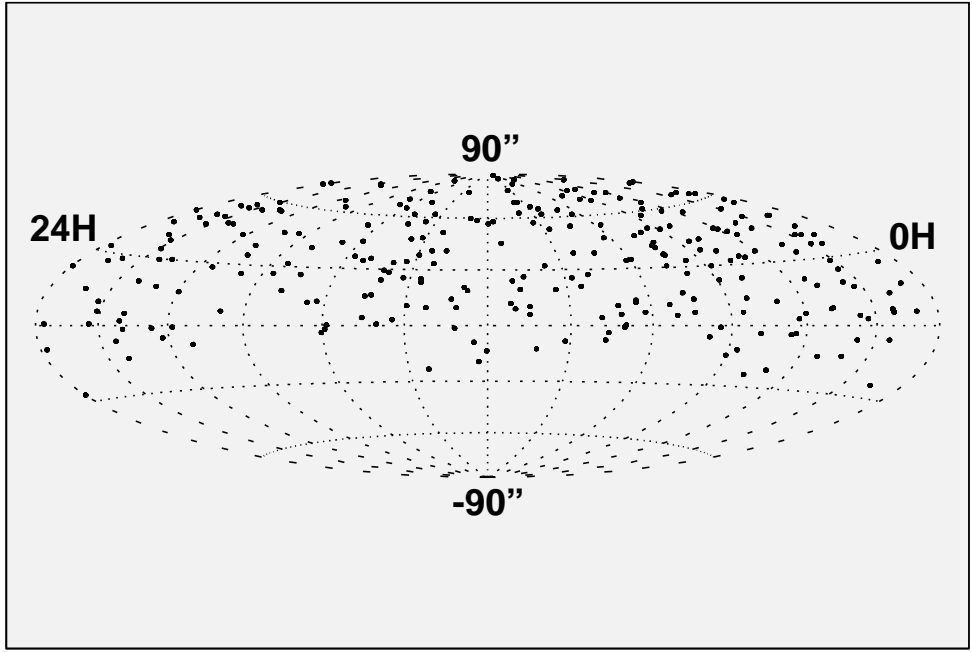}
    \caption{The 271 event stereo data set in Hammer-Aitoff projection.}
    \label{total}
  \end{figure}
\end{center}

\begin{figure}
  \includegraphics[angle=0.,width=\columnwidth]{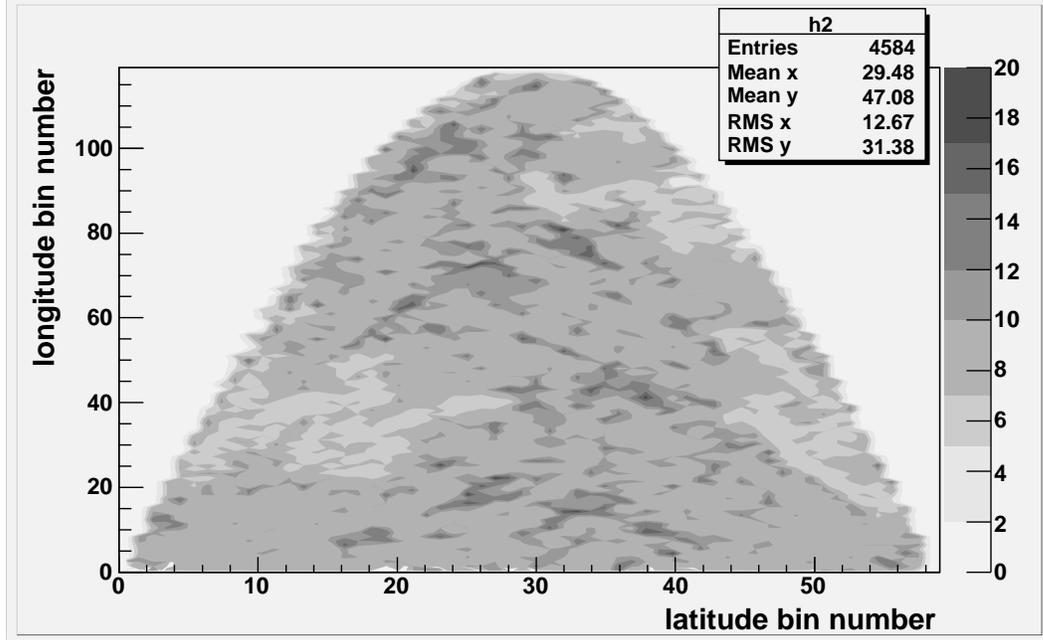}
  \caption{The bin occupancy values of normal great circles to the stereo data set . The sky was divided into 4584 bins of equal size (${6}^{\circ} \times {6}^{\circ}$ ) with the north pole at the zenith. The bins are arbitrarily numbered in latitude from north pole to south and around the axis.}
  \label{his}
\end{figure}

Figure~\ref{total} shows the particle's arrival directions (in equatorial coordinates) for the 271 events used in this analysis in Hammer-Aitoff projection. After applying the Hough Transform to the data set, the bin occupancy in dual-space is calculated. The resulting histogram is shown in Figure~\ref{his}.

\section{Search for significant excess.}
\label{sect-selection}

 In order to identify arcs, we look for apparent excesses over what is expected from an isotropic background. To estimate the significance of such excesses in each bin, we compared the data count to the distribution obtained from $2\times{10}^{4}$ isotropic simulated data sets.  Each simulated data set contains 271 events, and is based on a full detector simulation to account for the geometrical acceptance. In addition, the timing information is convolved to reproduce accurately the detector exposure in different directions of the sky over the detector live-time. The simulation used in this study is described in more details in ~\cite{r2}.

\begin{figure}
  \includegraphics[angle=0.,width=\columnwidth]{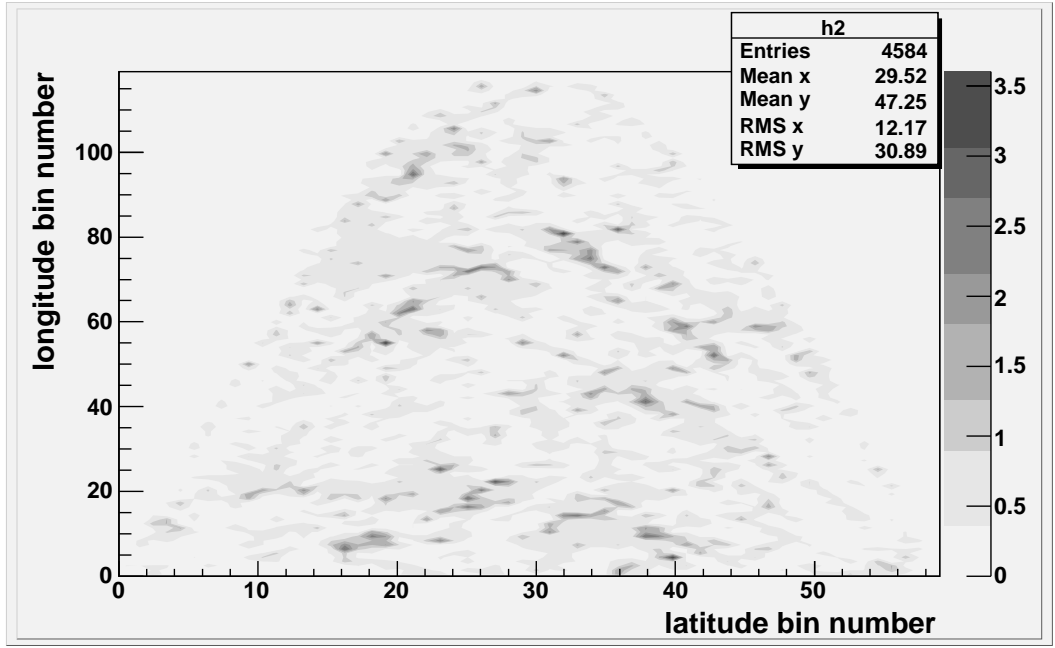}
  \caption{The significance values for each bin. As in figure~\ref{his}, the bins are indexed by both a longitudinal (RA) and a latitudinal (DEC) bin number.}
  \label{sig}
\end{figure}

From the Monte Carlo sets, we calculate the probability for each bin to have a simulated occupancy greater than or equal to that found in the data. We define the significance, S, for a bin to be given by $S = log(\frac{1}{P})$ and $P= \frac{M}{N}$, where N is the total number of simulated data sets, and M is the number of MC sets which recorded an occupancy greater or equal to that seen in the data bin. The resulting significance values are shown in  Figure~\ref{sig}, where each bin is indexed by both a longitudinal (RA) and a latitudinal (DEC) bin number. Bins with S $>2.5$ are designated as ``high significance''.  

\begin{figure}
  \includegraphics[angle=0.,width=\columnwidth]{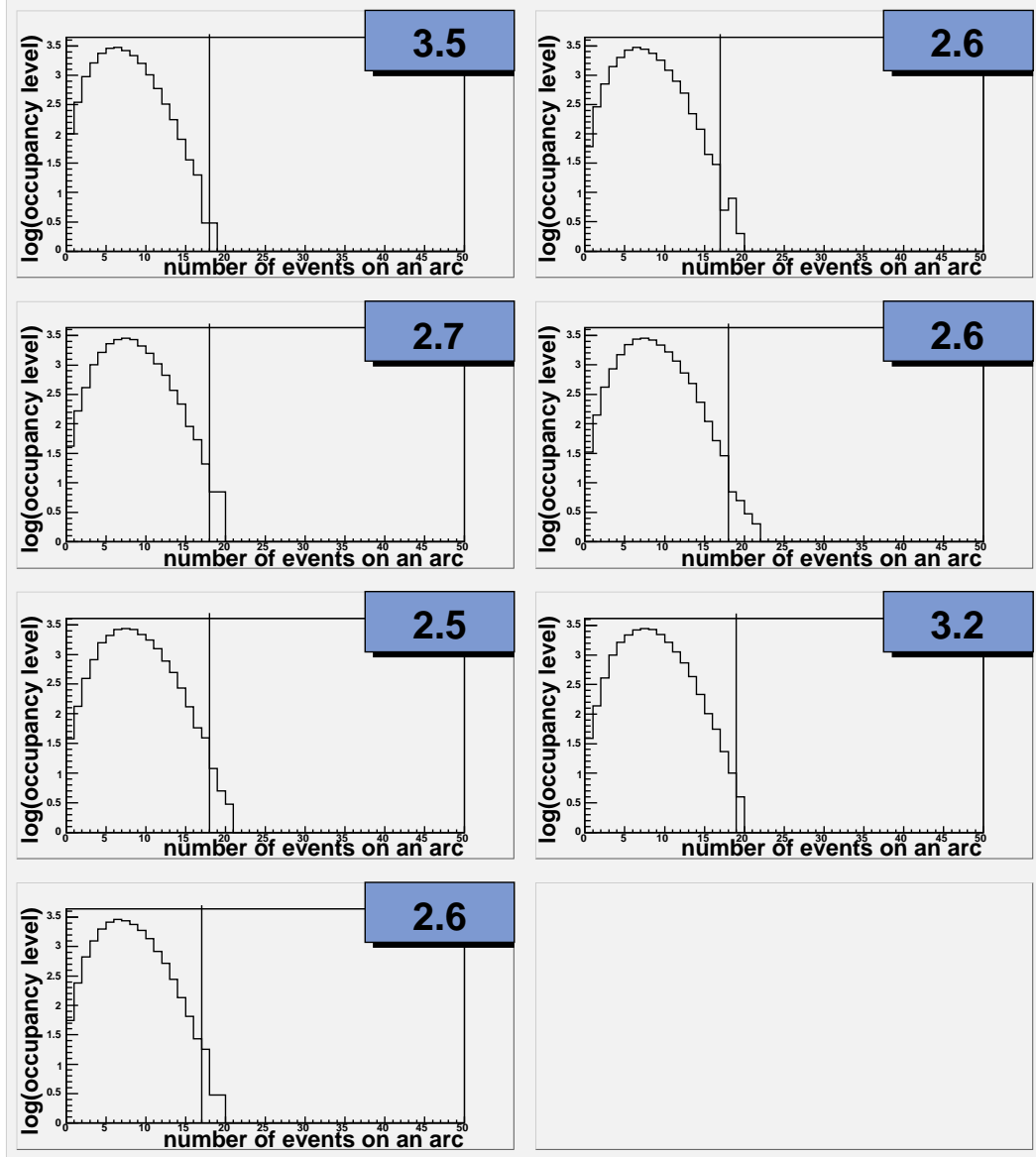}
  \caption{Histogram of bin x from $2\times{10}^{4}$ simulated data sets for the seven most significant (S $> 2.5$) bins in the data. The corresponding data bin value is indicated in each case by a vertical line. The S values are shown in the upper right hand corner.}
  \label{hisarcs}
\end{figure}

  For the seven high-significance bins we show the occupancy histogram for the $2 \times {10}^{4}$ isotropic MC set in Figure~\ref{hisarcs}, plotted in log scale. A vertical line is drawn on each plot at the bin occupancy level found in the data. The probability of randomly finding at least this occupancy level in an isotropic data set is then the ratio of the area under curve to the right of the line to the whole area ($2\times{10}^{4}$). The significance ``S'' of each of the seven bins is shown in the upper right hand corner of each histogram.

\section{Global Significance of data excess}
\label{sect-un-cal}
The next study we made was to determine whether the arcs found in the data are consistent with random fluctuations over the full sky, not just in the location in which they occurred. To do this, we compare our $2\times{10}^{4}$ isotropic simulated sets to an independent set of 1000 isotropically simulated data sets, each again with 271 events. We applied the same analysis and cuts as we did to the real stereo data set. Including the ``significance cut'' of S$>2.5$. We found that 99.6$\%$ of the time, the maximum significance of the isotropically generated data sets exceeded 2.5. Also, in 41.2$\%$ of the samples, the simulated data sets recorded a significance in at least one bin of $\geq 3.5$, which is the highest significance found in the real data set. Finally, the greatest number of events found on an arc in the real data is 19, as shown in Figure~\ref{hisarcs}. This was also found to be quite common in the simulation, since arcs with S$>2.5$ and at least 19 events were found in 90.4$\%$ of the simulated isotropic sets.

\section{Conclusion}

The Hough transform was applied to the HiRes stereo data and
seven arcs were found with significance $S > 2.5$. However, when a global significance analysis was performed, similar results were commonly reproduced within isotropic data sets of equivalent size. Hence, the observation of seven arcs in the data is not significant. We conclude that we do not see any evidence of arc-like structures in the HiRes stereo data set.

\section{Acknowledgments}
This work is supported by the National Science Foundation under contracts NSF-PHY-9321949, NSF-PHY-9322298, NSF-PHY-9974537, NSF-PHY-0071069, NSF-PHY-0098826, NSF-PHY-0140688,  NSF-PHY-0245328, NSF-PHY-0307098,  and NSF-PHY-0305516, as well as by Department of Energy grant FG03-92ER40732. We gratefully acknowledge the contribution from the technical staffs of our home institutions. We thank Gustavo Medina Tanco for providing Figure~\ref{OLENTO}. We gratefully acknowledge the contributions from the University of Utah Center for High Performance computing. The cooperation of Colonels E. Fisher, G. Harter, and G. Olsen, the US Army and the Dugway Proving Ground staff is appreciated.   

%
%
%

\end{document}